\documentclass[12pt]{article}
\usepackage{amssymb}
\font\teneuf=eufm10  \font\seveneuf=eufm7 \font\fiveeuf=eufm5
\newfam\euffam \def\gr{\fam\euffam\teneuf}

\textfont\euffam=\teneuf \scriptfont\euffam=\seveneuf
\scriptscriptfont\euffam=\fiveeuf
\def\be{\begin{equation}}
\def\ee{\end{equation}}
\def\bea{\begin{eqnarray}}
\def\eea{\end{eqnarray}}

\def\bprop{\begin{proposition}\rm}
\def\eprop{\end{proposition}}
\def\blemm{\begin{lemma}\small}
\def\elemm{\end{lemma}}
\def\bremark{\begin{remark}}
\def\eremark{\end{remark}}

\def \a {\alpha}
\def \b {\beta}

\def \g {\gamma}

\def \m  {\mu}

\def \z {\zeta}

\def \Lgl {\widetilde{\gr gl}}

\def \Lgl {\widetilde{\gr gl}}

\def \di {\partial}

\def \tr{{\rm tr}}
\def \res{{\rm res}}
\def \End{{\rm End}}

\def \otimescomma {\ {\scriptstyle {{}_\otimes\atop {{}^,}}}}
\def\nchi{\hbox{\raise 2.5pt\hbox{$\chi$}}}
\def\grgl{{\gr gl}}

\def\nchi{\hbox{\raise 2.5pt\hbox{$\chi$}}}
\def\BB{{\cal B}}

\def\II{{\cal I}}

\def\NN{{\cal N}}

\def\bfC{{\bf C}}

\def\bfI{{\bf I}}

\def\bfu{{\bf u}}
\def\bfv{{\bf v}}

\newtheorem{lemma}{Lemma}[section]
\newtheorem{remark}{Remark}[section]

\newtheorem{proposition}{Proposition}[section]

\begin{document}
\begin{flushright}
CRM-2908 (2002)\\
%nlin.SI/02xxxxx
\end{flushright}
\vspace{0.2cm}
\begin{center}
\begin{Large}
\textbf{Superintegrability, Lax matrices and separation of variables}
\footnote{ Work supported in part by the Natural Sciences and Engineering
Research Council of Canada (NSERC) and the Fonds FCAR du Qu\'ebec.}
\end{Large}\\
\vspace{1.0cm}
\begin{large} {J. Harnad }\footnote{harnad@crm.umontreal.ca} and
 { O. Yermolayeva}\footnote{yermolae@crm.umontreal.ca}
\end{large}
\\
\bigskip
\begin{small}
 {\em Centre de recherches math\'ematiques,
Universit\'e de Montr\'eal\\ C.~P.~6128, succ. centre ville, Montr\'eal,
Qu\'ebec, Canada H3C 3J7} \\
\smallskip
{\em Department of Mathematics and
Statistics, Concordia University\\ 7141 Sherbrooke W., Montr\'eal, Qu\'ebec,
Canada H4B 1R6} \\
\end{small}
\bigskip
\bigskip
%%%%%%%%%%%%%%%%%%%%%%%%%%%%%%%%%  Abstract %%%%%%%%%%%%%%%%%%%%%%%%%%%%%%%
{\bf Abstract}
\end{center}
\smallskip
\begin{small}
We show how the superintegrability of certain systems can be deduced from the
presence of multiple parameters in the rational Lax matrix representation. This
is also related to the fact that such systems admit a separation of variables
in parametric families of coordinate systems.

\end{small}
\bigskip
\bigskip
%%%%%  Section 1. Rational Lax matrix representations %%%%%%%%%%%%%%%%

\section{Rational Lax matrix representations of integrable systems}

\subsection{Classical $R$-matrix theory of commuting isospectral flows}

  In the classical $R$-matrix approach to finite dimensional integrable systems
\cite{STS, FT, H2}, there is a Poisson map from the phase space into a space of
$r \times r$ Lax matrices $\NN(\lambda)$ depending rationally, trigonometrically or
elliptically on a spectral parameter $\lambda$. The Poisson bracket
is defined by the relation
\be
\{\NN(\lambda) \otimescomma \NN(\mu)\} = [r(\lambda-\mu), \ \NN(\lambda) \otimes \bfI
   +\bfI \otimes \NN(\mu)] \ ,  \label{RmatrPB}
\ee
where both sides are interpreted as elements of  $\End(\bfC^r \otimes
\bfC^r)$. The symbol $\{ \ \otimescomma \ \}$ signifies a simultaneous tensor
product in $\End ({\bfC^r}) \sim \grgl(r)$  and  Poisson brackets in the
components, and $r(\lambda - \mu) $ denotes the classical $R-$matrix. The simplest
case is the rational $R$-matrix,
\be
r(\lambda ) :=  {P_{12} \over \lambda}, \qquad P_{12}(\bfu \otimes \bfv):= 
\bfv \otimes \bfu   \label{rPermut} \ ,
\ee
with  $\NN(\lambda)$ a rational function of $\lambda$.
\bea
\NN(\lambda) &=& \BB(\lambda) + \sum_{i=1}^n\sum_{a=1}^{n_i}
{N_{ia}\over (\lambda-\a_i)^i}
\\
\BB(\lambda) &:=& \sum_{i=1}^{n_0}B_i \lambda^i\ ,
\qquad \qquad N_{ia},
B_i \in \grgl(r) \ .
\eea
Equations (\ref{RmatrPB}), (\ref{rPermut})
define the standard linear, rational $R$-matrix structure.

   It follows from the properties of  classical $R$-matrices \cite{FT,
STS} that elements of the algebra of spectral invariants $\phi(\NN)\in
\II(\Lgl(r))$  Poisson commute amongst themselves and generate commuting
isospectral flows determined by  the Lax equations:
\be
\bf\dot{\NN}= \pm [(d\phi_\NN)_\pm, \NN]  \ ,
\ee
where $\NN$ is here thought of as an element of the   loop
algebra $\Lgl(r)$, identified in a standard way with its dual
${\Lgl}^*(r)$ through the trace-residue pairing, and  $(\ .\ )_{\pm}$ denotes
projection to the $\pm$ components relative to the usual splitting of the loop
algebra into positive and negative components
\be
\Lgl(r)) = \Lgl(r))_+ + \Lgl(r))_- \ 
\ee
(i.e. those admitting holomorphic continuations to the interior (+) and
exterior (-) of the unit circle respectively with the latter normalized to
vanish at $\infty$). The spectral invariants generate a maximal Poisson
commuting algebra on generic symplectic leaves, defining completely integrable
systems
\cite{H1, H2}; i.e., there are as many functionally independent generators as
half the dimension of the leaf. 

\subsection{$2\times 2$ rational Lax matrices}

    In the following, we shall limit our discussion to the case of $2\times 2$
Lax matrices, although most of the considerations that follow are easily
extended to higher rank. We may without loss of generality take $\NN(\lambda)$ to be
traceless (since the trace coefficients are Casimirs)
\be
\NN(\lambda) = \left (\begin {array}{cc} h(\lambda) & e(\lambda)\\ f(\lambda) & -h(\lambda) \end {array}
\right )  \ ,
\ee
where the rational functions $e(\lambda), f(\lambda), h(\lambda)$ satisfy the Poisson bracket
relations
\bea
\{h(\lambda), \ e(\mu)\} &=&{e(\lambda)-e(\mu) \over \lambda - \mu}  \cr
\{h(\lambda), \ f(\mu)\} &=&-{f(\lambda)-f(\mu) \over \lambda - \mu}   \cr
\{e(\lambda), \ f(\mu)\} &=&-2{h(\lambda)-h(\mu) \over \lambda - \mu} \ .  \label{sl2PB} 
\eea

 For this case, the ring $\II(\Lgl(2))$ of spectral
invariants, when restricted to the symplectic leaves of the $R$-matrix Poisson
structure, is generated by the quadratic trace invariants; i.e., the
coefficients determining the numerator of the rational function
\be
\Delta(\lambda):= - {1\over 2} \tr(\NN^2(\lambda) ) = h^2(\lambda) -{1\over 2} \left(e(\lambda)
f(\lambda) +f(\lambda) e(\lambda)\right) \ .
\ee
(The order in the last two terms is irrelevant of course, but it is written
here in a form that will also be valid in the quantum version below.)
If, for example, the polynomial part $\BB(\lambda)$ of $\NN(\lambda)$ is taken
to vanish, and only first order poles appear in $\NN(\lambda)$, we have
\bea
e(\lambda)&:=&  \sum_{i=1}^n {e_i\over \lambda-\a_i}   \cr
f(\lambda)&:=& \sum_{i=1}^n {f_i\over \lambda-\a_i}   \cr
h(\lambda)&:=&  \sum_{i=1}^n {h_i\over \lambda-\a_i},  \label{simplepoles}
\eea
where the quantities $\{e_i, f_i, h_i\}_{i=1\dots n}$ are a set of $n$
${\bf sl} (2)$ generators, which may be canonically coordinatized as:
\bea
e_i&:=& {1\over 2}\left(y_i^2 +
{\mu_i^2\over x_i^2}\right) \cr
f_i&:=& {1\over2}x_i^2     \cr
h_i&:=&{1\over 2}x_i{y_i} ,
\qquad i=1, \dots n,
\eea
where $\{\mu^2_i\}_{i=1\dots n}$  are the values of the  ${\bf sl} (2)$ Casimir
invariants and $\{x_i, y_i\}_{i=1\dots n}$ form a set of canonical coordinates
on the symplectic leaves .

\subsubsection{Parametric dependence of invariants and superintegrability}

Again, taking the case when the polynomial part $\BB(\lambda)$ of $\NN(\lambda)$
vanishes (but not necessarily just first order poles), a complete set of
generators is given by
\be
\phi_{ia} := \res_{\lambda=\a_i}(\lambda-\a_i)^a \tr(\NN^2(\lambda)), \qquad
i=1,\dots n, \ a=0, \dots n_i-1 \ .
\ee
These commute amongst themselves, but they each depend upon the pole locations
$\{\a_i\}_{i=1\dots n}$ in $\NN(\lambda)$. However, the linear combination:
\be
\phi_{SI}:= \sum_{i=1}^n \a_i \phi_{i0} = \res_{\lambda=\infty} \tr
(\NN^2(\lambda))
\ee
does not depend on the $\a_i$'s.  In general, there  is no reason for the
invariants $\phi_{ia}(\a_i)$ to commute with each other for different
choices of the $\a_i$'s. But, regardless of the values chosen, they will
commute with $\phi_{SI}$. Since the $\phi_{ia}(\a_i)$'s for different
choices of $\a_i$'s in general do not generate the same algebra of functions,
we may conclude that, taken together, for different evaluations of the
parameters $\{\a_i\}$, there are more functionally independent integrals that
Poisson commute with $\phi_{SI}$ than half the dimension of the symplectic leaf,
and hence the Hamiltonian system it generates is superintegrable. (In fact, in
most cases, it may be shown to be maximally superintegrable; see the examples
below.)

   In particular, if we take the case of purely simple poles as above in
(\ref{simplepoles}), the resulting Hamiltonian is:
\be
\phi_{SI} = {1\over 2} \sum_{i=1}^n x_i^2 \sum_{j=1}^n y_j^2
- {1\over 2} (\sum_{i=1}^n x_i y_i)^2
+ {1\over 2} \sum_{i=1}^n x_i^2  \sum_{i=1}^n {\mu_i^2 \over x_i^2} \ ,
\ee
which, when constrained to the (co)tangent bundle of the $n-1$ sphere $S^{n-1}$
\be
\sum_{i=1}^n x_i^2 = 1,  \qquad \sum_{i=1}^n x_iy_i =0 \ ,
\ee
yields the superintegrable system 
\be
h_{\rm Ros} = {1\over 2} \sum_{j=1}^n y_j^2
+ {1\over 2} \sum_{i=1}^n {\mu_i^2 \over x_i^2} \ ,
\ee
which is the trivial case of the Rosochatius system (without a harmonic
oscillator potential).

\subsubsection{Separation of variables}

Another viewpoint that helps to explain the superintegrability of systems
arising in this way is to note that they may be completely separated in a
canonical coordinate system determined by the values of the pole
parameters $\{\a_i\}$ which, for the ${\bf sl}(2)$ case with simple poles with
the phase space constrained to $S^{n-1}$ as above reduces to the sphero-conical
system
$\{\lambda_i,
\zeta_i\}_{i=1\dots n-1}$ defined by:
\be
\sum_{i=1}^n {x_i^2 \over \lambda - \a_i}= {\prod_{j=1}^{n-1}(\lambda-\lambda_j) \over
\prod_{i=1}^n(\lambda-\a_i)},  \qquad \z_i :={1\over 2} \sum_{i=1}^n {x_i y_i \over
(\lambda -\a_i)} \ .
\ee
These are just the points $(\lambda_i, \z_i)$ on the invariant spectral curve
\be
\z^2 + {1\over 2} \Delta (\lambda) =0 
\ee
where the matrix element $f(\lambda)$ vanishes and  $\z_i=
h(\lambda_i)$ are the eigenvalues at these points. These are particular cases of the
spectral {\it Darboux coordinates} of
\cite{H1, H2}. (Note that these become hyperellipsoidal coordinates if there is
a constant term added in the definition (\ref{simplepoles}) of $f(\lambda)$.)

   The point to note is that the separation of variables occurs in these
coordinates simultaneously for {\it all} the invariants $\phi_{ia}$, viewed as
generators of Hamiltonian flows. But again, since the leading term spectral
invariant $\phi_{SI}$ does not depend on the values of the parameters $\a_i$,
it admits a separation of variables in {\it any} of the family of
sphero-conical (or hyperellipsoidal) coordinates obtained by varying these
parameters. This simultaneous separability in multiple coordinates may  be
viewed as an alternative explanation of the origin of the superintegrability of
such systems. (In fact, both these viewpoints are a result of  the
classical $r$-matrix setting, and in a sense may be considered as equivalent.)

In the examples given below in the following section, the same principle
is used to deduce superintegrable systems from ${\bf sl}(2)$ Lax matrices
satisfying the Poisson bracket relations (\ref{RmatrPB}).

\subsubsection{Quantum integrable systems}

The above discussion is easily extended to the canonically quantized version
of such systems. All that must be done is to replace the matrix elements
defining $\NN(\lambda)$ by their quantized forms $\hat{e}(\lambda), \hat{f}(\lambda),
\hat{h}(\lambda)$, which must satisfy the commutator analogs of the Poisson bracket
relations (\ref{sl2PB})
\bea
[\hat{h}(\lambda), \ \hat{e}(\mu)] &=&{\hat{e}(\lambda)-\hat{e}(\mu) \over \lambda - \mu}
\cr
[\hat{h}(\lambda), \ \hat{f}(\mu)] &=&-{\hat{f}(\lambda)-\hat{f}(\mu) \over \lambda - \mu} 
\cr
[\hat{e}(\lambda), \ \hat{f}(\mu)] &=&-2{\hat{h}(\lambda)-\hat{h}(\mu) \over \lambda - \mu},
\label{sl2CR}  \ .
\eea
These can be realized by canonical quantization of the underlying classical
phase space variables. For example, in the case of simple poles only, with
vanishing polynomial term $\BB(\lambda)$, we have:
\bea
\hat{e}(\lambda)&:=&  \sum_{i=1}^n {\hat{e}_i\over \lambda-\a_i}   \cr
\hat{f}(\lambda)&:=& \sum_{i=1}^n {\hat{f}_i\over \lambda-\a_i}   \cr
\hat{h}(\lambda)&:=&  \sum_{i=1}^n {\hat{h}_i\over \lambda-\a_i},  \label{simplepolesQ}
\eea
where the ${\bf sl}(2)$ generators $\{\hat{e}_i, \hat{f}_i, \hat{h}_i\}$ may be
represented by the operators
\bea
\hat{e}_i&:=& {1\over 2}\left({\di^2\over \di x_i^2}-
{\mu_i^2\over x_i^2}\right) \cr
\hat{f}_i&:=& {1\over2}x_i^2     \cr
\hat{h}_i&:=&{1\over 2}\left(x_i{\di \over \di x_i} +{1\over 2}\right),
\qquad i=1, \dots n, \label{q-oper}
\eea
and the commuting invariants are similarly given by the coefficients of the
numerator polynomial of the quantum spectral invariant:
\be
\hat{\Delta}(\lambda):= \hat{h}^2(\lambda) -{1\over 2}
\left(\hat{e}(\lambda) \hat{f}(\lambda) + \hat{f}(\lambda) \hat{e}(\lambda)\right)  \ . \label{q-inv}
\ee

The resulting systems are similarly quantum integrable, and separable in the
same coordinates as the classical ones \cite{HW} and, for the same reasons as
above, the quantum version of the Hamiltonian $\phi_{SI}$ is superintegrable.

In the following section, a number of examples of such classical and quantum
superintegrable systems will be given.

\section{Examples of superintegrable classical and quantum systems}

The examples given below arise in the framework of the so-called Krall-Scheffer
problem \cite{KS} of describing all two-dimensional analogs of classical
orthogonal polynomials which result in nine classes of second-order partial
differential equations on the plane or on constant curvature surfaces. It was
shown in \cite{HVYZ},\cite{HYZ} that all nine cases are connected with
superintegrable systems. The following are some illustrative examples.

\subsection{Example 1.  The sphere.}

\subsubsection{Classical Lax Matrix}

The first case corresponds to three simple poles and vanishing $B(\lambda)$ . The Lax
matrix has the form:
\be
N(\lambda )=\frac{N_1}{(\lambda -\alpha )}+\frac{N_2}{(\lambda -\beta )}+\frac{N_3}{(\lambda -\gamma )}
%\begin{displaymath}
=\left(
\begin{array}{cc}
h(\lambda ) & f(\lambda ) \\
e(\lambda ) & -h(\lambda )
\end{array}
\right)
%\end{displaymath}
\ee
where the matrix elements of the  $N_i$ generate a Poisson bracket
realization of
$({\bf sl} (2))^3$ :
\be
\ {N_1}=\frac 12\left(
\begin{array}{cc}
s_1p_1 & p{_1}^2+\frac{\mu _1^2}{s_1^2} \\
-s{_1}^2 & -s_1p_1
\end{array}
\right)
\ee
\be
\ {N_2}=\frac 12\left(
\begin{array}{cc}
s_2p_2 & p{_2}^2+\frac{\mu _2^2}{s_2^2} \\
-s{_2}^2 & -s_2p_2
\end{array}
\right)
\ee
\be
\ {N_3}=\frac 12\left(
\begin{array}{cc}
s_3p_3 & p{_3}^2+\frac{\mu _3^2}{s_3^2} \\
-s{_3}^2 & -s_3p_3
\end{array}
\right)
\ee
Here $(p_1,p_2,p_3)$ are canonically conjugate to $(s_1,s_2,s_3)$ (and these
coincide with the coordinates $\{x_i, y_i\}_{i=1\dots n}$ above).

\subsubsection{Commuting invariants}

The invariants are the coefficients of:
\be
-\frac 12\tr N(\lambda )^2=\frac{H_1}{(\lambda -\alpha )}+\frac{H_2}{(\lambda
-\beta )}+\frac{H_3}{(\lambda -\gamma )}+\frac{{\mu _1}^2}{(\lambda -\alpha
)^2}+\frac{{\mu _2}^2}{(\lambda -\beta )^2}+\frac{{\mu _3}^2}{(\lambda
-\gamma )^2}  \ .
\ee

Note that only two of the integrals  $H_1$, $H_2$ and $H_3$ are independent,
since their sum is  zero. The Hamiltonian of the problem is given
by their linear combination:
\be
H=\alpha H_1 + \beta H_2 + 
\gamma H_3=\frac{1}{2}({p_1}^2+{p_2}^2+{p_3}^2) + \frac{{\mu_1}^2}{{s_1}^2}+
\frac{{
\mu_2}^2}{{s_2}^2}+ \frac{{\mu_3}^2}{{s_3}^2}  .
\ee
This describes the  Rosochatius system with harmonic oscillator terms absent
on the cotangent bundle of a two-sphere in $\mathbb R^3$:
\be
{s_1}^2+{s_2}^2+{s_3}^2=1,  \qquad s_1 p_1+s_2 p_2 + s_3 p_3=0 \ .
\ee \label{cotb}
The integrals $H_1$, $H_2$ and $H_3$ are as follows:

\bea
H_1 &=&-\frac{1}{2} \frac{{L_{13}}^2 + {{\mu_3}^2{s_1}^2}/{{s_3}^2}+
{{\mu_1}^2{ s_3}^2}/{{s_1}^2} }{\alpha - \gamma}- \frac{1}{2} \frac{{L_{12}}^2
+ {{\mu_1}^2{s_2}^2}/{{s_1}^2}+ {{\mu_2}^2{s_1}^2}/{{s_2}^2} }{\alpha - \beta}
\cr
H_2&=&-\frac{1}{2} \frac{{L_{23}}^2 + {{\mu_3}^2{s_2}^2}/{{s_3}^2}+ 
{{\mu_2}^2{
s_3}^2}/{{s_2}^2} }{\beta - \gamma}+ \frac{1}{2} \frac{{L_{12}}^2 + {{\mu_1}
^2{s_2}^2}/{{s_1}^2}+ {{\mu_2}^2{s_1}^2}/{{s_2}^2} }{\alpha - \beta}
\cr
H_3&=& \phantom{-}\frac{1}{2} \frac{{L_{23}}^2 + {{\mu_3}^2{s_2}^2}/{{s_3}^2}+ 
{{\mu_2}^2{ s_3}^2}/{{s_2}^2} }{\beta - \gamma}+ \frac{1}{2} \frac{{L_{13}}^2 +
{{\mu_3} ^2{s_1}^2}/{{s_3}^2}+ {{\mu_1}^2{s_3}^2}/{{s_1}^2} }{\alpha - \gamma}
\ ,
\eea
where $L_{ij}=s_1p_2-s_2p_1$ .

Note that  the Hamiltonian $H$ is independent of the parameters $
(\alpha, \beta, \gamma)$, whereas the invariants $H_1$, $H_2$  do depend on
them. Therefore, different choices for the parameters give distinct
integrals that commute with $H$, but do not commute with each other.

\subsubsection{Separating coordinates }

The  separating coordinates $(\lambda_1, \lambda_2)$  in this case are
sphero-conical coordinates. The corresponding momenta are denoted $(\xi_1,
\xi_2)$ . They are related to $(s_1,s_2,s_3)$ and $(p_1, p_2, p_3)$ by:
\be
{s_1}^2 = \frac
{(\alpha-\lambda_1)(\alpha-\lambda_2)}{(\alpha-\beta)(\alpha-\gamma)} \qquad \xi_1 = -\frac{1}{2} ( \frac{s_1p_1}{\lambda_1-\alpha} +
\frac{s_2p_2}{\lambda_1-\beta} + \frac{-s_1p_1 -s_2p_2}{\lambda_1-\gamma} )
\\
\ee
\be
{s_2}^2 = \frac
{(\beta-\lambda_1)(\beta-\lambda_2)}{(\beta-\alpha)(\beta-\gamma)} \qquad \xi_2 = -\frac{1}{2} ( \frac{s_1p_1}{\lambda_2-\alpha} +
\frac{s_2p_2}{\lambda_2-\beta} + \frac{-s_1p_1 -s_2p_2}{\lambda_2-\gamma} )
\ee
\be
{s_3}^2 = \frac
{(\gamma-\lambda_1)(\gamma-\lambda_2)}{(\gamma-\alpha)(\gamma-\beta)} \qquad 
\ee

\subsubsection{Quantum system }

The quantum versions of the integrals above, 
denoted  $\hat {H_1},\hat {H_2}, \hat {H_3}$ , are obtained by replacing
the matrix elements of $N(\lambda)$ by the corresponding differential
operators,
$\hat{e}(\lambda), \hat{f}(\lambda), \hat{h}(\lambda)$, which in the case of simple poles  are
as in  (\ref{simplepolesQ})-(\ref{q-inv}).

The quantization procedure leads to replacing the $L_{ij}$'s by their quantum
version:
\be
\hat L_{ij}=\sqrt{-1} (s_i{\partial}/{\partial}{s_j}-s_j{\partial}/{\partial}
{s_i})
\ee
Introducing the functions
\be
\omega_{jk}^2:={{\mu_j}^2{s_k}^2}/{{s_j}^2}+{{\mu_k}^2{s_j}^2}/{{s_k}^2} \qquad
j,k=1..3
\ee
and denoting $\alpha=\alpha_1 ~,~ \beta=\alpha_2 ~,~ \gamma=\alpha_3$,
we can present the quantum integrals as
\be
\hat {H_i}= - \frac{1}{2} \sum_{k \not= i} { \frac {\hat {L}_{ik} + \omega_{ik}^2 } {\alpha_i - \alpha_k} } \qquad i,k=1..3
\ee
The quantum Hamiltonian is
\be
\hat H=-\frac{1}{2} ({{\partial}_1}^2+{{\partial}_2}^2+{{\partial}_3}^2)  +\frac{{\mu_1}^2}{{s_1}^2}+
\frac{{
\mu_2}^2}{{s_2}^2}+ \frac{{\mu_3}^2}{{s_3}^2} . 
\ee

The separating coordinates are the configuration
space part of the ones for the classical case $(\lambda_1,\lambda_2)$.

\subsection{Example 2.  The hyperboloid.}

\subsubsection{Classical Lax Matrix}

Consider now a Lax matrix with one first order and one second order 
pole:

\be
N(\lambda )=\frac{N_1}{(\lambda -\alpha )}+\frac{N_2}{(\lambda -\alpha )^2}+
\frac{N_3}{(\lambda -\beta )},
\ee
with
\bea
{\ N_1}&=&\frac 12\left(
\begin{array}{cc}
s_1p_1+s_2p_2 & 2p_1p_2+ 2\g_1\g_2\\
-2s_1s_2 & -s_1p_1-s_2p_2
\end{array}
\right)
\cr
\ {N_2}&=&\frac 12\left(
\begin{array}{cc}
-s_2p_1 & -{p_1}^2-\g_2^2 \\
s{_2}^2 & s_2p_1
\end{array}
\right)
\cr
\ {N_3}&=&\frac 12\left(
\begin{array}{cc}
s_3p_3 & p{_3}^2+ \g_3^2\\
-s{_3}^2 & -s_3p_3
\end{array}
\right)
\eea
Here we have introduced the following notations 
\be
2\g_1\g_2:=\frac {2 \mu_2^2
s_1}{s_2^3} - \frac {2 \mu_1
\mu_2}{s_2^2}~, ~\g_2^2:=-\frac{{\mu_2}^2}{s_2^2}~,~\g_3^2:=\frac{\mu
_3^2}{s_3^2}.
\ee

The matrix elements of $(N_1,N_2)$ generate a Poisson bracket realization of
the jet extension 
${\bf sl}(2)^{(1)*}$ while those of $N_3$ generate a second  ${\bf sl} (2)$.

\subsubsection{Commuting invariants}

The trace formula again gives us only two
independent commuting invariants $H_1$ and $H_2$
\be
-\frac 12\tr N(\lambda )^2=\frac{H_1}{(\lambda -\alpha )}+\frac{H_2}{(\lambda
-\alpha )^2}-\frac{\mu _1\mu _2}{(\lambda -\alpha )^3}+\frac{{\mu _2}^2}{
2(\lambda -\alpha )^4}+\frac{H_3}{(\lambda -\beta )}-\frac{{\mu _3}^2}{
2(\lambda -\beta )^2}
\ee
since, by taking the residue we obtain
\be
H_1+H_3=0  \ .
\ee
The superintegrable Hamiltonian in this case is:
\be
H=(\alpha -\beta )H_1+H_2-\frac 12{\mu _3}^2 = 2p_1p_2  - {p_1}^2 + p{_3}^2 + 2\g_1\g_2 -\g_2^2 + \g_3^2.
\ee
The quadratic constraint now  defines a hyperboloid 
\be
2s_1s_2+{s_3}^2=1 \ .
\ee
In the ambient  coordinates the integrals $H_1$ and $H_2$ are
\bea
H_1&=& \frac{ (s_1p_3-s_3p_2)(s_3p_1-s_2p_3) - \g_3^2 s_1 s_2 - 2 \g_1\g_2
s_3^2}{ \alpha - \beta }  - \frac{((s_3p_1-s_2p_3)^2 + \g_3^2 s_2^2 + \g_2^2
s_3^2 }{2(\alpha - \beta)^2}
\cr
H_2&=&\frac{1}{2}(s_1p_1-s_2p_2)^2 - 2 \g_1 \g_2 s_1 + \frac {(s_3p_1-s_2p_3)^2
+ \g_3^2 s_2^2 + \g_2^2 s_3^2 }{2(\alpha -\beta)}.
\eea

Again, whereas the Hamiltonian $H$ does not depend on the parameters $
(\alpha, \beta)$ the integrals $H_1,~H_2$ do, which provides an
explanation for the superintegrability in this case.

\subsubsection{Separating coordinates }
These are determined by the relations:
\be
{s_3}^2 = \frac
{(\beta-\lambda_1)(\beta-\lambda_2)}{(\alpha-\beta)^2} \qquad  {\xi_1} = -\frac{1}{2} (
\frac{s_1p_1+s_2p_2}{\lambda_1-\a} - \frac{s_2p_1}{(\lambda_1-\a)^2} + \frac{s_3p_3}{\lambda_1-\beta} )
\ee 

\be
{s_2}^2 = - \frac
{(\alpha-\lambda_1)(\alpha-\lambda_2)}{(\alpha-\beta)} \qquad {\xi_2} = -\frac{1}{2} (
\frac{s_1p_1+s_2p_2}{\lambda_2-\a} - \frac{s_2p_1}{(\lambda_2-\a)^2} + \frac{s_3p_3}{\lambda_2-\beta} )
\ee

\be
{s_1s_2} = - \frac{1}{2} ( \frac
{(\beta-\lambda_1)(\beta-\lambda_2)}{(\alpha-\beta)^2} - 1) .
\ee

\subsubsection{Quantum system}

The quantized integrals $\hat{H_1},\hat{H_2},\hat{H_3}$ are obtained as
before by replacing all conjugate variables by the corresponding differential
operators. The quantum integrals may then be expressed as
\bea
\hat {H_1}&=& -\frac{(s_1{\partial}_3-s_3{\partial}_
2) (s_3{\partial}_1-s_2{\partial}_
3) + \g_3^2 s_1 s_2 + 2 \g_1\g_2 s_3^2}{\alpha - \beta } + \frac{
(s_3{\partial}_1-s_2{\partial}_ 3)^2 -\g_3^2 s_2^2 - \g_2^2 s_3^2 }{2(\alpha -
\beta)^2}
\cr
\hat {H_2} &=& \frac{1}{2}\hat {L}_{12}^2-2 \g_1 \g_2 s_1 - \frac {
(s_3{\partial}_{s_1}-s_2{\partial}_ {s_3})^2 - \g_3^2 s_2^2 - \g_2^2 s_3^2}
{2(\alpha -\beta)} ,
\eea
where ${\partial}_k:={\partial}/{\partial} {s_k}$.

The quantum Hamiltonian is
\be
\hat H=2{\partial} _1{\partial}_2  - {{\partial}_1}^2 + {{\partial}_3}^2 + 2\g_1\g_2 -\g_2^2 +
\g_3^2,
\ee
and this again separates in the configuration space coordinates $(\lambda_1,\lambda_2)$.

\subsection{Example 3. The plane.}

\subsubsection{Classical Lax Matrix}

For the cases with zero curvature like the example to follow, the polynomial
part
$\BB(\lambda)$ of the Lax matrix does not vanish.
The simplest case involves two distinct finite poles in $N(\lambda )$ and
constant $B(\lambda)$
\[
N(\lambda )=\left(
\begin{tabular}{ll}
$0$ & $-a$ \\
$1$ & $0$
\end{tabular}
\right) +\frac 1{2(\lambda -\alpha )}\left(
\begin{array}{cc}
s_1p_1 & {p_1}^2 +\frac{\m_1^2}{s_1^2} \\
-{s_1}^2 & -s_1p_1
\end{array}
\right)
\]
\be
+\frac 1{2(\lambda -\beta )}\left(
\begin{array}{cc}
s_2p_2 & {p_2}^2 +\frac{\m_2^2}{s_2^2}\\
-{s_2}^2 & -s_2p_2
\end{array}
\right) \ .
\ee

The matrix elements of the residues $N_1, \ N_2$ generate two copies of ${\bf
sl}(2)$.

\subsubsection{Commuting invariants}

The invariants of motion are defined by:
\be
-\frac 12\tr N(\lambda )^2=\frac{H_1}{(\lambda -\alpha )}+\frac{H_2}{(\lambda
-\beta )} + \frac{{\mu _1}^2}{(\lambda -\alpha
)^2}+\frac{{\mu _2}^2}{(\lambda -\beta )^2} - a .
\ee

The superintegrable Hamiltonian in this case is given by
\be
H=\frac 14\res_\infty \tr N(\lambda )^2 =\frac
14({p_1}^2+{p_2}^2+a(s{_1}^2+s{_2}^2)+\frac{{\mu _1}^2}{{s_1}^2}+
\frac{{\mu _2}^2}{{s_2}^2}) ,
\ee
which gives an isotropic oscillator together
with Rosochatius terms. As before $(p_1,p_2)$ are canonically conjugate to $
(s_1,s_2).$

In terms of the ambient space coordinates the integrals $H_1$ and $H_2$ are :

\bea
H_1&=& p_1^2 +a s_1^2 + \frac{\m_1^2}{s_1^2} - \frac{1}{2(\a-\b)}
(L_{12}^2+\frac{\m_1^2 s_2^2}{s_1^2}+\frac{\m_2^2 s_1^2}{s_2^2})
\cr
H_2&=&p_2^2 + a s_2^2 - \frac{\m_2^2}{s_2^2} + \frac{1}{2(\a-\b)} (L_{12}^2 +
\frac{\m_1^2 s_2^2}{s_1^2}+ \frac{\m_2^2 s_1^2}{s_2^2} ) .
\eea
Where $L_{12}:=s_1p_2-s_2p_1$ and $H= \frac{1}{4}(H_1+H_2)$. Here the additional
integral results from the parametric dependence on $(\a-\b)$.

\subsubsection{Separating coordinates }

The separating coordinates $(\lambda _1,\lambda _2, \xi_1, \xi_2)$ in
this case are defined by
\be
{s_1}^2= 2 \frac{(\beta -\lambda _1)(\beta -\lambda _2)}{(\alpha -\beta)} \qquad \xi_1 = - \frac{1}{2} ( \frac{s_1p_1}{\lambda_1 -\alpha} 
+ \frac{s_2p_2}{\lambda_1 -\beta} )
\ee
\be
{s_2}^2= - 2 \frac{(\alpha -\lambda _1)(\alpha -\lambda _2)}{(\alpha - \beta)} \qquad \xi_2 = - \frac{1}{2} ( \frac{s_1p_1}{\lambda_2 -\alpha} 
+ \frac{s_2p_2}{\lambda_2 -\beta} )
\ee

\subsubsection{Quantum system}

The Hamiltonian of the corresponding quantum problem is 
\be
\hat H=\frac 14\res_\infty \tr \hat N (\lambda )^2 =\frac
14({{\partial}_1}^2 + {{\partial}_2}^2 + a(s{_1}^2 + s{_2}^2) + \frac{{\mu
_1}^2}{{s_1}^2} +
\frac{{\mu _2}^2}{{s_2}^2}) = \frac{1}{4}(\hat H_1 + \hat H_2).
\ee
The quantum integrals $\hat H_1$  and $\hat H_2$ are:

\bea
\hat H_1&=& {\partial}_1^2 + a s_1^2 + \frac{\m_1^2}{s_1^2} - \frac{1}{2(\a-\b)}
(\hat L_{12}^2 + \frac{\m_1^2 s_2^2}{s_1^2} +
\frac{\m_2^2 s_1^2}{s_2^2} )
\cr
\hat H_2&=&{\partial}_2^2 + a s_2^2 - \frac{\m_2^2}{s_2^2} + \frac{1}{2(\a-\b)}
(\hat L_{12}^2 + \frac{\m_1^2 s_2^2}{s_1^2} +
\frac{\m_2^2 s_1^2}{s_2^2} )
\eea
and the separating coordinates are again $(\lambda_1,\lambda_2)$, which depend on the
additional parameter $(\a-\b)$.

%%%%%%%%%%%%%%%%%%%%%%%%%%  Bibliography  %%%%%%%%%%%%%%%%%%%%%%%%%%%%%%%%%%

\end{document}